# Charge Density Refinement of the Si (111) 7x7 surface

J. Ciston[1*], A. Subramanian[1], I.K. Robinson[2], L.D. Marks[1]

[1] Department of Materials Science and Engineering, Northwestern University, USA

[2] London Centre for Nanotechnology and Department of Physics and Astronomy, University College London, UK

## Abstract

We report an experimental refinement of the local charge density at the Si (111) 7x7 surface utilizing a combination of x-ray and high energy electron diffraction. By perturbing about a bond-centered pseudoatom model, we find experimentally that the adatoms are in an anti-bonding state with the atoms directly below. We are also able to experimentally refine a charge transfer of 0.26±0.04 $e^-$ from each adatom site to the underlying layers. These results are compared with a full-potential all-electron density functional DFT calculation.

       61.05.cp  (X-ray diffraction)

       61.05.jm  (CBED, selected-area electron diffraction, nanodiffraction)

       73.20.-r  (Electron states at surfaces and interfaces)

Two of the most powerful techniques for determining structures are x-ray diffraction and transmission electron microscopy or diffraction. It is very well established that in the bulk these techniques can be used not just to determine atomic positions, but going beyond this to measure local charge-density variations [1-3]. At any surface one of the most important scientific issues is the local redistribution of electron density, not simply the atomic positions, since the electron density governs most of the properties and plays a central role in many properties of scientific and technological importance.

Perhaps the most interesting and challenging mono-species surface structure is the Si (111) 7x7 reconstruction first observed by Schiller and Farnsworth[4] and finally solved decades later by Takayanagi et al [5, 6] who proposed the well-known dimer, adatom, stacking fault (DAS) structure. The large size of this structure has provided a challenge to density functional theory (DFT) calculations with the first *ab-initio* relaxations and surface energy computations utilizing LDA [7] pseudopotential methods not appearing until 1992 [8, 9]. The first GGA [10] pseudopotential DFT calculations of the DAS structure did not appear until 2005 and is qualitatively consistent with the earlier LDA results. This structure has also played an important role in the development of scanning probe techniques, both in their infancy [11], and in pushing the limits past atomic resolution in AFM measurements of bond energies [12] to sub-atomic resolution in STM studies where the out of plane adatom orbitals were resolved distinctly [13, 14].

In this letter, we present a surface diffraction refinement of site-specific charge transfer at the Si (111) 7x7 surface, which pushes the limit of the amount of information a combination of x-ray and transmission high energy electron diffraction data can provide. By perturbing about a bond-centered pseudoatom model, we find experimentally that the adatoms are in an anti-bonding state with the atoms directly below which may finally explain the anomalous height of the adatom above the surface [15]. We are also able to experimentally refine a charge transfer of 0.26±0.04 $e^-$ from each adatom site to the underlying layers. These experimental results are compared with all-electron full-potential DFT structural refinements.

The X-ray measurements were conducted at X16A beamline at the National Synchrotron Light Source at the Brookhaven National Labs. We used a 6x30x0.2 mm$^3$ Si(111) wafer slice, which was first etched with HF to regrow a controlled oxide layer. Then it was flashed by passage of current to 1200C for 5 seconds, cooled very quickly to about 900C then slowly to 750C, spanning the phase transition region. The pressure in the chamber during the measurements lasting 84 hours was around $5\times10^{-10}$ torr. 1054 symmetry-reduced structure factors were measured by numerical integration of rocking scans about each point in the reciprocal space, wide enough to allow a full background subtraction, then corrected for Lorenz factor, polarization, and active area [18].

For the electron diffraction measurements, undoped Si (111) single crystal samples were cut into 3mm discs and mechanically dimpled then thinned to electron transparency by a HF and $HNO_3$ chemical etch. These samples were then transferred into a UHV chamber with a base pressure of $8\times10^{-11}$ torr and annealed by electron bombardment for 20 min at 720 C to produce the 7x7 reconstruction. Samples were transferred *in-situ* to a Hitachi UHV-H9000 transmission electron microscope for off-zone-axis parallel nanobeam diffraction experiments. A total of 3540 in-plane measurements were reduced to 77 p3m1 symmetry unique reflections (p6mm Patterson symmetry) using a Tukey-biweight method to a resolution of 0.65 Å$^{-1}$.

For the refinement, isolated atomic form factors (IAM) were treated according to the expansion of Su and Coppens[19]; these were also transformed to electron scattering factors utilizing the Mott-Bethe formula. The global bonding charge density was refined using a bond-centered pseudo-atom (BCPA) formalism which treats the 1s, 2s, and 2p core electrons identically to the IAM model, applies a fixed modified Slater orbital expansion for the 3s and 3p valence states, and utilizes distorted bulk parameterization of the Si-Si bond density with Gaussian charge clouds as described in [16, 17]. By utilizing a BCPA model to parameterize the bond charges as a function of only the bond length, one may refine the global surface charge density without the addition of adjustable parameters to the refinement

Diffraction refinements were performed utilizing a robust degree of freedom reduced $\chi$ figure of merit scaled for intensity conservation. A modified robust Hamilton R-test was used to assess the statistical significance of adding adjustable parameters to the fit. For the combined x-ray and electron case, there are 1131 independent data points and 129 adjustable parameters: 114 for non-symmetry constrained positions of 49 atoms, 4 temperature factors (only the adatoms were treated with separate in- and out-of-plane terms), and 11 scaling terms yielding 1002 degrees of freedom (DoF). The refined model consisted of 61 atoms representing the addition of one 12-atom layer constrained to bulk positions at the bottom surface. No preferential weighting was given to the electron dataset and all data points were weighted according to the inverse of their errors. Although the electron data is in principle more sensitive to bonding because small perturbations to bonding electrons induce large changes in the screening of the core potential, the electron dataset is not large enough to be used alone due to the large number of adjustable parameters required to fully describe the structure. It is important to note that the diffraction refinement does not utilize any information from the DFT structural relaxation presented later.

The first step toward refining site-specific changes to the bond density is to apply the global BCPA model to the system as a whole and determine if indeed the diffraction data is sensitive to bonding information. We find nothing to contradict the stability of the DAS structure. Applying the BCPA model to the refinement yields a reduction in $\chi$ from an IAM value of 2.689 to 2.599 for the bonded case. This reduction in $\chi$ is statistically significant to over 99.999% due to the large number of degrees of freedom. If only the x-ray dataset is utilized, the improvement in $\chi$ due to the BCPA bonding approximation is similar. The RMS deviation of the IAM-refined atomic positions from the BCPA refined positions is 0.035Å. As expected, the inclusion of subtle bonding effects has relatively little impact on the atomic positions as x-ray scattering is dominated by the core electrons. CIF files of the Si (111)-7x7 structures refined to the diffraction dataset and DFT relaxation are available online in the EPAPS repository.

Once the global correction to the valence charge density has been applied, it is possible to probe more subtle site-specific perturbations about the BCPA charge density. The first feature that will be examined is the nature of the bond between the adatoms and the atomic sites directly below. To probe this, we performed refinements for four distinct cases:

a) adatoms are 3-fold coordinated and not bonded to the third layer
b) adatoms are 4-fold coordinated and bonded via a BCPA feature to the third layer
c) adatoms are 4-fold coordinated and exhibit an anti-bonding state with the third layer atoms via a BCPA feature with the back-bond length, but opposite sign – essentially adding a "dangling bond" above the adatoms.
d) All the adatoms, rest atoms and the hole atom have a dangling bond, where the effective distance for the rest atoms and hole atom was taken as the mean of that for the adatoms.

In each of these cases, all of the other Si-Si pairs were treated under the BCPA formalism. We find from the diffraction refinement that case c, an anti-bonded adatom, is the most favorable to a 96% confidence level. This configuration of the adatom orbitals is consistent with prior AFM[12] and STM[13, 14] studies of this surface. While based solely upon the refinement we cannot fully justify adding a dangling bond to the rest atoms and hole atoms (c and d are indistinguishable), physically it is more reasonable to assume that they are chemically similar. The adatom dangling bonds of case c are interpreted as an anti-bonding state due to the close proximity (2.85Å) to the atoms directly below (see Figure 1). Conversely, the rest atoms and corner hole atoms are 4Å distant from the atoms below, so this is interpreted as merely a dangling unpaired electron rather than an anti-bond. Case b is the worst-performing in the diffraction refinement due to the unstable 5-fold coordination of the third layer back-bond. Refinements of the various back-bonding states become degenerate if the electron diffraction portion of the data is removed from the refinement, which is not surprising; adding a dangling bond delocalizes the electron density which will increase the electrostatic potential, even in projection, to which the TED data is very sensitive.

We will next turn to the refinement of a feature with a predominantly in-plane component: charge transfer from the adatom sites to/from the underlying tripod atoms (see Figure 1). If the charge transfer at all four adatom sites is constrained to be identical, the value is refined to be 0.26±0.04 $e^-$ per adatom. Applying the Hamilton R-test to determine if the addition of this adjustable parameter is allowed yields a confidence value of 99%. Although the adatoms represent only 12 of the 249 atoms in the refined structure, the data is extremely sensitive to charge defects fractions of an electron in magnitude. An attempt was made to refine the adatom charge transfer against the x-ray data alone, but this refinement proved unstable.

Another unusual feature of the DAS structure are the buried dimer atoms which have previously been shown to exhibit bond lengths 6±2% longer than bulk values [20], indicating a slightly weaker bond compared to the bulk value. We have refined the value of the dimer bond density against the diffraction data to be 0.37±0.04 $e^-$ (92% of the bulk value) by allowing the charge clouds within the BCPA model to vary in magnitude with all dimers treated identically. The refined magnitude of the dimer bond is invariant of the application of charge transfer to the adatom bond indicating that these two site-specific parameters are independent variables. Although the dimer bond refinement is stable, the addition of the adjustable parameter for dimer bond strength fails the Hamilton test with a confidence value of <40% so it is suggestive, rather than definitive. Refining the dimer strength against only the x-ray data set yields a significant reduction in $\chi$, but the value of the dimer bond diverges and produces values of 5-7 $e^-$ which is unphysical.

We also performed a DFT structural relaxation using the all electron Wien2k code[21] with the exchange correlation contribution approximated using the PBE96 GGA functional[10] as well as the more accurate TPSS functional [22], the latter used only for a correction to the total energy after a GGA relaxation. Prior experience has also shown that the absolute energy error in a PBE surface calculations can be approximated by the difference in the total energy between the PBE and TPSS functionals [23]. DFT surface energies may be found in Table 1.

The unit cell used was 26.882 Å x 26.882 Å x 28.220 Å (a 1.25% volume expansion relative to the experimental unit cell as determined by a total energy volume optimization for bulk Si). The surface slab was made centrosymmetric and comprised of 12 layers, 61 independent atoms with P-3m1 symmetry (498 total atoms), and 10Å of vacuum between surfaces. All atoms were free to move during the relaxation including the central unreconstructed layers. Technical parameters for the calculation were Si muffin-tin radii of 2.12, an RKMAX of 6.75, and a single k-point at the special point (5/18,1/9,0). The structure was relaxed until all forces acting on the atoms were under 0.2 eV/Å. All calculations were performed spin-unpolarized as the spin-zero closed shell state has previously been calculated to be the ground state for the 7x7-DAS structure[24]. A more detailed analysis of the DFT results can be found elsewhere [25].

The RMS in-plane deviation of experimentally refined atomic positions with respect to the relaxed DFT values is 0.08Å. However, the out of plane performance is not as good with an RMS deviation from DFT values of 0.27Å mostly due to excess outward relaxation of the surface layers in the diffraction refinement. This reflects the fact that the in-plane electron diffraction data has much smaller errors than the comparable x-ray datasets, and also that the in plane x-ray data is of higher quality than the out of plane rod scans. The out of plane uncertainty is also exhibited in the refined temperature factors of the adatom layer with B values of 15 $Å^2$ in the out of plane direction and 0.92 $Å^2$ in plane. A portion of this quite large value is due to an accumulation of collective out of plane vibrations from the underlying layers.

To asses the magnitude of charge transfer to/from the adatom sites, the total DFT electron density was integrated over each atomic basin utilizing the Bader atoms in molecules[26] (AIM) approach coded into the Wien2k package. The average charge transfer determined from the DFT calculations is 0.16±0.03 $e^-$ per adatom, similar to the experimentally refined value of 0.26±0.04 $e^-$. Note that the diffraction refinement is able to address the charge at each atomic site by transferring density directly from the spherical component of the 3sp shell of one atom to that of another, whereas the AIM analysis is a method of partitioning the global charge density of the structure to individual atoms. While the AIM method can be somewhat misleading in assigning electrons from the total charge density as "belonging" to particular atoms, it allows us to qualitatively verify that the magnitude and direction of the diffraction-refined charge transfer is reasonable. AIM analysis may also be used to determine the "strength" of a bond by looking at the charge density at the bond critical points. We find the density of the dimer bond is 93% of the bulk value. This is in remarkable agreement to the value of 92% refined against the diffraction data.

A (110) slice through the charge density of the 7x7 surface cell allows for visualization of all of the symmetry-inequivalent adatoms, rest atoms, and corner holes. A plot of the difference between the full DFT charge density and the charge density of superpositioned isolated atoms is shown in Figure 1a and compared with in 1b a difference map just using the BCPA model with no dangling bonds and in 1c with dangling bonds on the adatoms, rest atoms, hole atom and the charge transfer described above. We are not refining against the DFT data, but we will argue that the fact that Figures 1a and 1c are qualitatively much more similar than Figures 1a and 1b supports a contention that we are refining here physically significant features in the density, not overfitting noisy data. A closer inspection of the DFT difference density at the adatoms reveals that the adatom charge density is qualitatively similar to a Si-Si anti-bond as indicated by the wedge of off-axis excess charge. This confirms the aforementioned finding of the diffraction refinement which favored an anti-bonded adatom backbond.

In this study we have been able to refine the first three dimensional site-specific surface charge density from diffraction data. The stable refinement of an anti-bonded adatom may explain the anomalous length of this bond. In addition charge transfer from the adatom site to the underlying tripod is directly refined to the diffraction data. Although the x-ray diffraction data is shown to be globally sensitive to bulk valence charge density effects, these experiments appear insufficiently sensitive to refine site-specific perturbations to the bonding at the surface without the addition of data from electron diffraction. Newly designed SXRD setups with greater emphasis on stability and more efficient detectors might allow even larger and more accurate data sets that could allow these further issues to be addressed. However, electron diffraction is confirmed by our analysis to be more sensitive to local bonding effects and should be the method of choice for studying surface charge density.

**Acknowledgements**

This work was supported by the NSF on grant number DMR-0455371/001 (JC and LDM). X-ray diffraction data was collected by A. Ghosh at the X16A beamline of the National Synchrotron Light Source at Brookhaven National Laboratory. Use of the National Synchrotron Light Source, Brookhaven National Laboratory, was supported by the U.S. Department of Energy, Office of Science, Office of Basic Energy Sciences, under Contract No. DE-AC02-98CH10886.

**Figures**

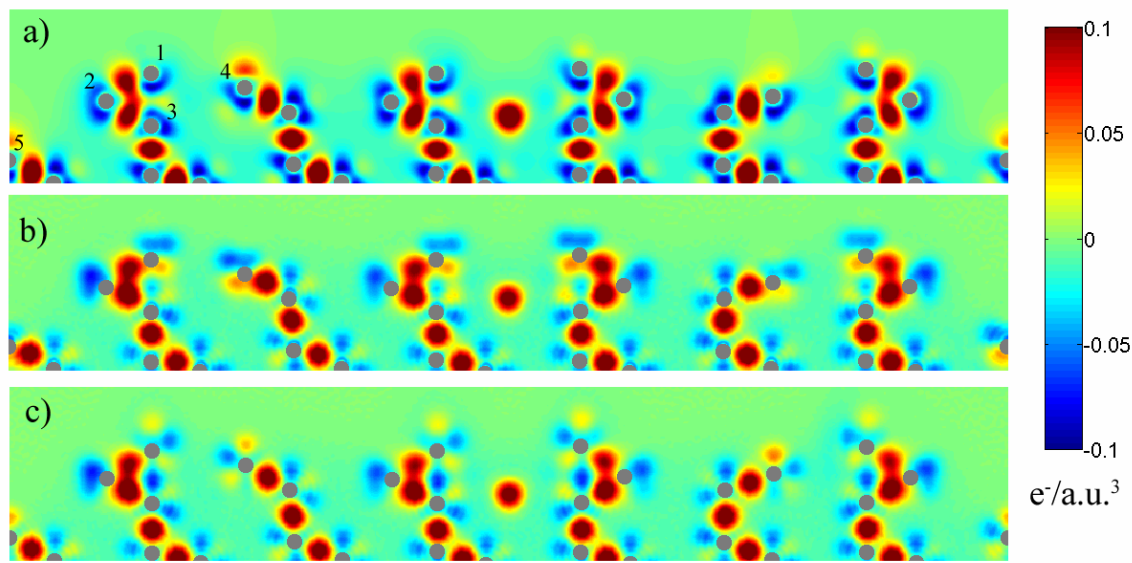

Figure 1: (110) slice through 7x7 unit cell (color online) of a) the DFT difference density, b) the difference density using just the conventional BCPA and c) a map of the charge density features fitted in the diffraction refinement including both the dangling bond features as well as the charge transfer. Silicon atom positions are shown in gray. Color scale is electrons per cubic atomic unit. Atoms labeled as follows: 1) adatom, 2) tripod atom, 3) backbond atom, 4) rest atom, 5) corner hole atom.

## Tables

| Method | eV / 1x1 |
|---|---|
| PBE full potential | 0.954 |
| TPSS full potential | 0.949 |
| PBE pseudopotential [24] | 1.044 |
| LDA pseudopotential [9] | 1.153 |

Table 1: DFT Surface energies per 1x1 unit cell